\documentstyle[12pt,emlines2]{article}
\def\bea{\begin{eqnarray}}
\def\eea{\end{eqnarray}}

\def\beq{\begin{equation}}
\def\eeq{\end{equation}}
\def\ba{\beq\new\begin{array}{c}}
\def\ea{\end{array}\eeq}

\makeatletter
\newdimen\normalarrayskip              
\newdimen\minarrayskip                 
\normalarrayskip\baselineskip
\minarrayskip\jot
\newif\ifold             \oldtrue            \def\new{\oldfalse}
\def\arraymode{\ifold\relax\else\displaystyle\fi} 
\def\eqnumphantom{\phantom{(\theequation)}}     
\def\@arrayskip{\ifold\baselineskip\z@\lineskip\z@
     \else
     \baselineskip\minarrayskip\lineskip2\minarrayskip\fi}
\def\@arrayclassz{\ifcase \@lastchclass \@acolampacol \or
\@ampacol \or \or \or \@addamp \or
   \@acolampacol \or \@firstampfalse \@acol \fi
\edef\@preamble{\@preamble
  \ifcase \@chnum
     \hfil$\relax\arraymode\@sharp$\hfil
     \or $\relax\arraymode\@sharp$\hfil
     \or \hfil$\relax\arraymode\@sharp$\fi}}
\def\@array[#1]#2{\setbox\@arstrutbox=\hbox{\vrule
     height\arraystretch \ht\strutbox
     depth\arraystretch \dp\strutbox
     width\z@}\@mkpream{#2}\edef\@preamble{\halign
\noexpand\@halignto
\bgroup \tabskip\z@ \@arstrut \@preamble \tabskip\z@ \cr}%
\let\@startpbox\@@startpbox \let\@endpbox\@@endpbox
  \if #1t\vtop \else \if#1b\vbox \else \vcenter \fi\fi
  \bgroup \let\par\relax
  \let\@sharp##\let\protect\relax
  \@arrayskip\@preamble}
\def\eqnarray{\stepcounter{equation}%
              \let\@currentlabel=\theequation
              \global\@eqnswtrue
              \global\@eqcnt\z@
              \tabskip\@centering
              \let\\=\@eqncr
              $$%
 \halign to \displaywidth\bgroup
    \eqnumphantom\@eqnsel\hskip\@centering
    $\displaystyle \tabskip\z@ {##}$%
    \global\@eqcnt\@ne \hskip 2\arraycolsep
         $\displaystyle\arraymode{##}$\hfil
    \global\@eqcnt\tw@ \hskip 2\arraycolsep
         $\displaystyle\tabskip\z@{##}$\hfil
         \tabskip\@centering
    &{##}\tabskip\z@\cr}

\begin{document}

\begin{flushright}
ITEP/TH-21/00\\
hepth/0005066
\end{flushright}
\vspace{0.5cm}
\begin{center}
{\LARGE \bf Tunneling into the Randall-Sundrum Brane World}


\bigskip
{\bf A. Gorsky, K. Selivanov}

\vspace{0.5cm}
\bigskip
{ ITEP, Moscow, 117259, B.Cheryomushkinskaya 25}
\end{center}

\setcounter{footnote}0

\begin{abstract}
We suggest a modification of the Randall-Sundrum scenario which
doesn't involve branes with the negative tensions. In our scenario
four-dimensional brane world is produced by the external field.
The probability of this process is calculated and  the physical
features of the model are discussed.
\end{abstract}

1. The alternative to compactification suggested recently in \cite{rs}
attracts a lot of
attention nowadays. It seems to be one of the most promising approaches
to unification of gravity with other  interactions. It is actually
the development of the old idea of the brane world \cite{rubshap}
with additional localization of the gravity on the brane due to the
AdS geometry. The simplest version suggested in the original paper
deals with the $R^4\times \frac{S^1}{Z_2}$ geometry with the
$AdS_5$ metric. In the furthercoming
papers several modifications
have been investigated.

In particular,  it was suggested to consider
the noncompact fifth dimension \cite{grs} with the flat metric
at $|z|>z_0$, where z is the fifth coordinate. This scenario involves
three branes located along z and to provide vanishing of the
total cosmological constant
in four dimensions the sum of the brane tensions should vanish.
Further considerations  of the models of this type can be found
in \cite{more}.
Other  modification  \cite{kogan}   involves
three branes which
however locates on the compact fifth coordinate.  Among other
developments let us mention the possibility to have our world
on the brane which intersects with other branes in higher
dimensions \cite{intersec}. The nonstationary versions of the RS
scenario also have been analyzed \cite{nonstat}
and the possibility to have
the reasonable inflation have been outlined \cite{inflation}.
Recently there was attempt to obtain the reasonable
4d world from the flat fifth dimension \cite{dvali}.

Most of scenarios suffer from the necessity
to consider at least one brane
with the negative tension. The negative
tension branes are hard to imagine to be  the stable objects.
This quite unnatural ingredient
definitely gets down the attractiveness of the models.
The goal of this
letter is to find the model which would keep at least some pleasant
features of the RS scenario
and its modifications but doesn't
involve the negative tension branes.

The model suggested in this letter has the model of Gregory, Sibiryakov
and Rubakov (GRS) as the closest relative from the RS family.
However unlike  GRS model  in our case there are no branes
with the negative tension as well as the jump of the cosmological
constant.
Two key ingredients of our model are the constant strength tensor for
the four-form field and the presence of the three brane junctions.

Perhaps,  the external four-form field may seem artificial.
Notice,  however,  that  many compactifications of the
M theory naturally involve fluxes of the four-form field
as, for instance, it was discussed recently in \cite{pol}.
The external field was used in \cite{bg} to avoid fine tuning
of brane tension/cosmological constant.
Notice also, that the external field is a sort of hidden in the GRS
construction since their cosmological constant is not a constant,
it jumps on the branes.

What concerns to the junctions, they  are very
common  in the string theory context as well. In general
one has to distinguish two different types of brane junctions.
The junctions of the first type involve the  NS or
RR D branes in the string theory and the junction itself
doesn't introduce the additional energy.
It the second situation branes appear as solitonic
solutions to the equations of motion in some field theories.
In SUSY case junctions of such effective branes
keep 1/4 of the initial SUSY and saturate the
specific central charges which are certainly known in Wess-Zumino
model \cite{wz} as well as SYM theory \cite{sym}. These central
charges provide the stability of the junctions of the effective branes
and amounts to the additional energy of the configuration.
One more possible source for the junctions are the
domain walls in SUGRA considered in RS context in \cite{sugra}.
In what
follows we will imply that there is no contribution to the
total energy from the junction manifold or it is much smaller then
the energies of the branes themselves, though we believe that
this assumption is not crucial,  our considerations should go
through with nonzero junction energy as well.

To clarify the role of the external field let us remind that
it can produce  brane pairs in the spontaneous Schwinger
type  tunneling
process \cite{bt}. We will need however more involved process
of the tunneling in the presence of the external branes. The
process of the induced tunneling that is the tunneling with branes in
initial state has been discussed in \cite{gs}. Here we will
exploit the "inverse" situation when additional brane appears in
the final state (see figure ).
It is  this brane we are presumably living on. We refer to it below
as RS-brane. Two others which play the
role of regulators in the four dimensional theory
will be referred to as
R-branes.  R-branes are charged with respect to the external field, RS-brane
is neutral.

Remarkably,  Einstein equations can be explicitely solved for
the quite complicated looking brane configuration on figure.
The metric consists of two AdS pieces sewed with each other
on the RS-brane  and with outside flat metric on the R-branes.
RS-brane inherits AdS metric, which however can be done as flat
as one wishes.
Of course, the metric of the "Euclidean" classical solution
determines
the ``Minkowski'' evolution of the Universe. Importantly,
the R-branes are not static, they accelerate as they should
in the external field.  RS brane remains at rest.

The paper is organized as follows. First we consider the simplified 1+1
example and calculate the probability of the creation of the
one-dimensional world. This simple calculation reveals the ideas
used but doesn't take into account the effects of gravity.
Then we consider the realistic case in higher dimensions and
show that there is the  "Euclidean" classical configuration (bounce)
which corresponds to the creation of the pair of branes with the
additional brane at rest in the final state. We
investigate the geometry
of the arising brane world and show that the four dimensional
cosmological constant can be made appropriately small. Finally
we discuss the physical features of the model and formulate some
open questions.

2. In this section we consider the simple example of the similar
process in two dimensions. The problem can be formulated as follows.
Electric field can create fermion-antifermion pair in  an
exponentially suppressed process. Suppose that the final state
consists of the fermion pair and an
additional bound state.

It is well known that the
spontaneous pair creation is described by the effective action
\beq
\label{action}
S=-eEA+TL
\eeq
where $L$ is the length of the  world-line of the particles produced
(as usual, particle-antiparticle history looks like  a closed world-line),
and  $A$ is the area surrounded by the world-line.
$T$ stands for  mass of the particle,  e stands for its charge ,  $E$ is
the external electric field.

Notice, by the way, that
upon  appropriate identification
of the parameters, the same effective action describes
false vacuum decay in (1+1) scalar field theory.  Particles are substituted
by kinks, electric field - by energy difference between false and true vacuum.

Extremal world-line for the action Eq.(\ref{action}) is a circle
of the radius $R=\frac{m}{E}$
and therefore the probability P of the spontaneous process looks as
follows
\beq
P \propto exp(-\frac{\pi T^2}{eE}).
\eeq

Generalization of the effective action
for the process with the additional particle
in the final state also allows purely geometrical consideration,  in the way
suggested in \cite{sv} for the induced vacuum decay in (1+1).
One considers effective action of the type of
\beq
\label{action2}
S=-eEA+T_{R}L_{R}+ T_{RS}L_{RS}
\eeq
where  $T_{R}$ stands for a mass of charge particles (analogues of  R-branes
in what follows), $L_{R}$ stands for length of their world-line,
$T_{RS}$ stands for a mass of the additional neutral particle
(analog of RS-brane),  $L_{RS}$ - for length of its world-line.
The geometry of the bounce  is presented
below \\[10pt]

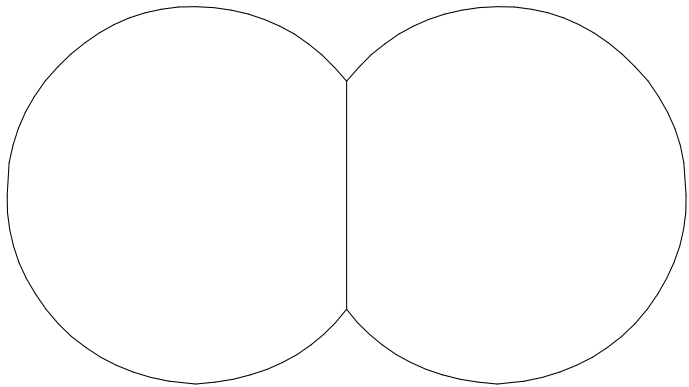
\special{em:linewidth 0.4pt}
\unitlength 1.00mm
\linethickness{0.4pt}
\begin{picture}(95.42,55.23)(-20.00,00.00)
\emline{50.08}{45.11}{1}{48.48}{46.96}{2}
\emline{48.48}{46.96}{3}{46.79}{48.62}{4}
\emline{46.79}{48.62}{5}{45.01}{50.09}{6}
\emline{45.01}{50.09}{7}{43.14}{51.37}{8}
\emline{43.14}{51.37}{9}{41.18}{52.46}{10}
\emline{41.18}{52.46}{11}{39.13}{53.35}{12}
\emline{39.13}{53.35}{13}{36.99}{54.06}{14}
\emline{36.99}{54.06}{15}{34.76}{54.57}{16}
\emline{34.76}{54.57}{17}{32.45}{54.90}{18}
\emline{32.45}{54.90}{19}{30.04}{55.03}{20}
\emline{30.04}{55.03}{21}{27.74}{54.97}{22}
\emline{27.74}{54.97}{23}{25.50}{54.70}{24}
\emline{25.50}{54.70}{25}{23.34}{54.22}{26}
\emline{23.34}{54.22}{27}{21.23}{53.54}{28}
\emline{21.23}{53.54}{29}{19.20}{52.65}{30}
\emline{19.20}{52.65}{31}{17.22}{51.55}{32}
\emline{17.22}{51.55}{33}{15.32}{50.25}{34}
\emline{15.32}{50.25}{35}{13.48}{48.75}{36}
\emline{13.48}{48.75}{37}{11.71}{47.03}{38}
\emline{11.71}{47.03}{39}{10.00}{45.11}{40}
\emline{10.00}{45.11}{41}{8.57}{43.07}{42}
\emline{8.57}{43.07}{43}{7.38}{40.97}{44}
\emline{7.38}{40.97}{45}{6.43}{38.80}{46}
\emline{6.43}{38.80}{47}{5.71}{36.56}{48}
\emline{5.71}{36.56}{49}{5.23}{34.25}{50}
\emline{5.23}{34.25}{51}{4.95}{29.93}{52}
\emline{4.95}{29.93}{53}{5.01}{27.62}{54}
\emline{5.01}{27.62}{55}{5.30}{25.36}{56}
\emline{5.30}{25.36}{57}{5.81}{23.14}{58}
\emline{5.81}{23.14}{59}{6.55}{20.98}{60}
\emline{6.55}{20.98}{61}{7.52}{18.86}{62}
\emline{7.52}{18.86}{63}{8.71}{16.80}{64}
\emline{8.71}{16.80}{65}{10.12}{14.78}{66}
\emline{10.12}{14.78}{67}{11.76}{12.81}{68}
\emline{11.76}{12.81}{69}{13.53}{11.15}{70}
\emline{13.53}{11.15}{71}{15.40}{9.68}{72}
\emline{15.40}{9.68}{73}{17.39}{8.40}{74}
\emline{17.39}{8.40}{75}{19.49}{7.31}{76}
\emline{19.49}{7.31}{77}{21.70}{6.42}{78}
\emline{21.70}{6.42}{79}{24.02}{5.72}{80}
\emline{24.02}{5.72}{81}{26.45}{5.22}{82}
\emline{26.45}{5.22}{83}{30.04}{4.83}{84}
\emline{30.04}{4.83}{85}{32.59}{5.07}{86}
\emline{32.59}{5.07}{87}{35.01}{5.48}{88}
\emline{35.01}{5.48}{89}{37.32}{6.05}{90}
\emline{37.32}{6.05}{91}{39.51}{6.79}{92}
\emline{39.51}{6.79}{93}{41.57}{7.70}{94}
\emline{41.57}{7.70}{95}{43.51}{8.78}{96}
\emline{43.51}{8.78}{97}{45.34}{10.02}{98}
\emline{45.34}{10.02}{99}{47.04}{11.43}{100}
\emline{47.04}{11.43}{101}{48.62}{13.01}{102}
\emline{48.62}{13.01}{103}{50.08}{14.76}{104}
\emline{50.08}{45.11}{105}{51.68}{46.96}{106}
\emline{51.68}{46.96}{107}{53.37}{48.62}{108}
\emline{53.37}{48.62}{109}{55.16}{50.09}{110}
\emline{55.16}{50.09}{111}{57.03}{51.37}{112}
\emline{57.03}{51.37}{113}{58.99}{52.46}{114}
\emline{58.99}{52.46}{115}{61.03}{53.35}{116}
\emline{61.03}{53.35}{117}{63.17}{54.06}{118}
\emline{63.17}{54.06}{119}{65.40}{54.57}{120}
\emline{65.40}{54.57}{121}{67.71}{54.90}{122}
\emline{67.71}{54.90}{123}{70.12}{55.03}{124}
\emline{70.12}{55.03}{125}{72.42}{54.97}{126}
\emline{72.42}{54.97}{127}{74.66}{54.70}{128}
\emline{74.66}{54.70}{129}{76.83}{54.22}{130}
\emline{76.83}{54.22}{131}{78.93}{53.54}{132}
\emline{78.93}{53.54}{133}{80.97}{52.65}{134}
\emline{80.97}{52.65}{135}{82.94}{51.55}{136}
\emline{82.94}{51.55}{137}{84.85}{50.25}{138}
\emline{84.85}{50.25}{139}{86.68}{48.75}{140}
\emline{86.68}{48.75}{141}{88.45}{47.03}{142}
\emline{88.45}{47.03}{143}{90.16}{45.11}{144}
\emline{90.16}{45.11}{145}{91.59}{43.07}{146}
\emline{91.59}{43.07}{147}{92.78}{40.97}{148}
\emline{92.78}{40.97}{149}{93.74}{38.80}{150}
\emline{93.74}{38.80}{151}{94.46}{36.56}{152}
\emline{94.46}{36.56}{153}{94.94}{34.25}{154}
\emline{94.94}{34.25}{155}{95.22}{29.93}{156}
\emline{95.22}{29.93}{157}{95.16}{27.62}{158}
\emline{95.16}{27.62}{159}{94.87}{25.36}{160}
\emline{94.87}{25.36}{161}{94.36}{23.14}{162}
\emline{94.36}{23.14}{163}{93.62}{20.98}{164}
\emline{93.62}{20.98}{165}{92.65}{18.86}{166}
\emline{92.65}{18.86}{167}{91.46}{16.80}{168}
\emline{91.46}{16.80}{169}{90.05}{14.78}{170}
\emline{90.05}{14.78}{171}{88.41}{12.81}{172}
\emline{88.41}{12.81}{173}{86.64}{11.15}{174}
\emline{86.64}{11.15}{175}{84.76}{9.68}{176}
\emline{84.76}{9.68}{177}{82.77}{8.40}{178}
\emline{82.77}{8.40}{179}{80.67}{7.31}{180}
\emline{80.67}{7.31}{181}{78.46}{6.42}{182}
\emline{78.46}{6.42}{183}{76.14}{5.72}{184}
\emline{76.14}{5.72}{185}{73.71}{5.22}{186}
\emline{73.71}{5.22}{187}{70.12}{4.83}{188}
\emline{70.12}{4.83}{189}{67.57}{5.07}{190}
\emline{67.57}{5.07}{191}{65.15}{5.48}{192}
\emline{65.15}{5.48}{193}{62.84}{6.05}{194}
\emline{62.84}{6.05}{195}{60.66}{6.79}{196}
\emline{60.66}{6.79}{197}{58.60}{7.70}{198}
\emline{58.60}{7.70}{199}{56.65}{8.78}{200}
\emline{56.65}{8.78}{201}{54.83}{10.02}{202}
\emline{54.83}{10.02}{203}{53.12}{11.43}{204}
\emline{53.12}{11.43}{205}{51.54}{13.01}{206}
\emline{51.54}{13.01}{207}{50.08}{14.76}{208}
\emline{50.08}{45.11}{209}{50.08}{14.76}{210}
\end{picture}

Euclidean time is assumed to go upward on the picture.
Trajectories of charged particles are segments of the circles of the same radius as in
spontaneous decay.  The world-line of the RS-particle is a straight line.
The angle at the junction is fixed by the ``force'' balance condition
$2T_{R}cos \alpha = T_{RS}$.
It is evident from the condition
above that the  particles with masses larger then 2T can not
be produced via this mechanism.
Notice also that details of interaction of the R and RS particles belong
to preexponential factor.

Now the calculation of the
probability $ P \propto exp(-S)$ is straightforward and yields
\beq
S=\frac{2T_{R}^2}{eE}(\pi - arcsin(1- \frac{T_{RS}^2}{4T_{R}^2})^{1/2}) +
\frac{2T_{R} T_{RS}}{eE}(1- \frac{T_{RS}^2}{4T_{R}^2})^{1/2}
\eeq
Remark that $t=0$ is the turning point of the bounce.  This is the critical
configuration created in Minkowski space. After that point in Minkowski space
R-particles move in the opposite directions while the RS-particle remains at
rest. Actually this is the model of the creation of the one-dimensional
Universe on the RS-particle.  That is this picture which will be generalized
in what follows for the higher-dimensional case. The new feature to be taken
into account are effects of gravity.

3. Higher dimensional bounce  is more or less obvious generalization
of  the two- dimensional one.  Namely we have
two symmetric segments of the spheres intersecting over the
junction manifold and the flat brane inside which has junction
manifold as a boundary. In what follows
to keep closeness to the original
GRS scenario we shall concentrate on the case of
three branes in five dimensions however consideration in the higher
dimensions is analogous.

The effective action we will work with will be
\beq
\label{action3}
S= \frac{1}{2k}\int d^5 x \sqrt g (-R + 2\Lambda)
+\frac{1}{k} \oint \sqrt g_{ind}K+
\frac{1}{2} \int d^5 x \sqrt g h^2 +
T_{R} \oint_{R} \sqrt g_{ind} +
T_{RS} \oint_{RS} \sqrt g_{ind}
\eeq
Let us explain ingredients in Eq.(\ref{action3}).
$h$ is Hodge dual scalar of the field strength $H=dB$ of the four-form
field, $H_{\mu_1\ldots \mu_5}=\sqrt g \epsilon_{\mu_1 \ldots \mu_5} h$.
Notice that in 5d this field does not propagate.  Field equations
say that in empty space $h$ is a constant, in our case it only jumps
at the charged R-branes by its charge e,
\beq
h_{+}- h_{-}=e
\eeq
where $h_{+}$ is outside value of $h$, $h_{-}$ is its inside value.
In the effective action Eq.(\ref{action3}) it is assumed that
field equations for $h$ are resolved (see \cite{bt}, where the spontaneous
brane creation without RS-brane was thoroughly studied, for more explanations
on this point). Notice, that after resolution, the charge of the R-branes
enters effective action only via $h^2$-term.

Constant k in Eq.(\ref{action3}) is the five dimensional gravitational
constant, $R$ - scalar curvature, $\Lambda$ stands for cosmological constant,
for which we assume that it is negative and exactly compensates
energy density of the field $h$ outside,
\beq
\Lambda + \frac{k h_{+}^2}{2} =0
\eeq
so that outside the metric is flat.  Notice right away that effective
cosmological constant  inside is
\beq
\label{cc}
\Lambda_{eff}=\Lambda + \frac{k h_{-}^2}{2}=
- \frac{k h_{+}^2}{2}+ \frac{k h_{-}^2}{2}.
\eeq
and hence the scalar curvature $R$ of the AdS metric inside
any of the two segments in figure reads
\beq
\label{curvature}
R=\frac{2d}{d-2} \Lambda_{-}=\frac{10}{3} \Lambda_{-},
\eeq
and the corresponding AdS radius, $R_{AdS}$, reads
\beq
\label {rads}
R^{2}_{AdS}=-\frac{2(d-1)(d-2)}{\Lambda_{eff}}=-\frac{24}{\Lambda_{eff}}.
\eeq

The origin of two of the three surface terms in Eq.(\ref{action3}) is obvious -
these are tension terms for R- and RS-branes. The third term,
$ \frac{1}{k} \oint \sqrt g_{ind}K$ is introduced to ensure
that variation of the curvature term does not depend on normal derivatives
of variation of the metric on the branes \cite{israel}.
Here $K=g_{ind}^{ij}K_{ij}$ stands for the trace of the external curvature
$K_{ij}$ of the branes, the integral is over all branes, every brane
contributing twice, with $K$ computed in metric on one or the other side of
the brane.

So much about ingredients in the effective action Eq.(\ref{action3}).
Let us now explain details of the bounce solution.
The metric to the right of the RS-brane inside R-brane   reads
\beq
\label{plus}
ds^{2}=\frac{dz^2+d\rho^2+\rho^2d\Omega^2_{3}}
{(1-\frac{(z-a)^2+\rho^2}{R^{2}_{AdS}})^2}
\eeq
while the metric to the left of the RS-brane inside R-brane   reads
\beq
\label{minus}
ds^{2}=\frac{dz^2+d\rho^2+\rho^2d\Omega^2_{3}}
{(1-\frac{(z+a)^2+\rho^2}{R^{2}_{AdS}})^2}
\eeq
where $z$ is coordinate along the axes of symmetry orthogonal to
RS-brane, $\rho$ is the radial coordinate in orthogonal to $z$
directions,  $d\Omega^2_{3}$ is the metric of the corresponding
3-sphere,  $a$ is a parameter.

The metrics Eq.(\ref{plus}),(\ref{minus}) are to be sewed
on R-branes with flat metric  and on RS-brane with
each other, in the sence that metrics themselves are continues,
while their normal derivatives jump according to the Israel
condition:
\beq
\label{Israel}
\Delta K_{ij}=\frac{kT}{d-1}g_{ij}=\frac{kT}{4}g_{ij}.
\eeq
On R-branes this condition fixes radius ${\bar R}$
of the spherical segments,
\beq
\label{rbounce}
{\bar R}=\frac{2T_{R}(d-1)}{h^{2}_{+}-h^{2}_{-}},
\eeq
which is of course the same as for bounce without
RS-brane,
{\footnote{We would like to warn the reader that we use different
coordinates from those in \cite{bt} and \cite{cdl}}}
and on RS-brane this condition fixes the parameter
$a$:
\beq
\label{a}
a=\frac{T_{RS}(d-1)}{h^{2}_{+}-h^{2}_{-}}
\eeq
Importantly,
\beq
\label{angle}
cos{\alpha}=\frac{a}{\bar R}=\frac{T_{RS}}{2T_{R}},
\eeq
which is precisely the ``force'' balance condition at the junctions.

Substituting these data into the effective action Eq.(\ref{action3})
one straigtforwardly obtains exponential factor for the probability
P of the process. It ranges between
\beq
P\propto e^{-S_{bt}}
\eeq
for very light RS-brane,  $T_{RS}\ll T_{R}$,  where
$S_{bt}$ is the action for the bounce without RS-brane, which has been
computed in \cite{cdl}, \cite{bt} (see also \cite{bkt}),
and
\beq
P\propto e^{-2S_{bt}}
\eeq
for  $T_{RS}=2 T_{R}$. More heavy RS-brane cannot be produced in this way.

Since the internal brane is located at $z=0$  the
induced metric of 4D world is immediately seen from Eq.(\ref{plus}).
It appears to be the
AdS space  with the AdS radius
\beq
R_{AdS4}^2=R_{AdS}^2-a^2
\eeq
and, correspondingly, the cosmological constant
\beq
\label{eff}
\Lambda_4=-\frac{12}{(R_{AdS}^2 -a^2)}
\eeq

4.  The key question in any model is if it reproduces the
standard gravity in four dimensions. Since our model is
the generalization of GRS one let us remind the key features
of GRS model \cite{grs, more}. Unlike the RS scenario graviton in
GRS model is quasi-localized since the zero mode of the
gravitational fluctuation is nonnormalizable. Therefore
four dimensional graviton can be considered as the long living
resonant state. It is clear that the scale which governs the
lifetime of the graviton is defined by the distance between
branes with positive and negative tensions. To get the
usual Newton law one has to integrate over some range in
the continuum spectrum but it is valid only at the intermediate
distances. The Newton gravity  holds in the region
$R_{AdS}\ll r \ll z_0$ and
gets modified outside this interval.

It may seem that we have no such region since
the tunneling takes place when $R_{AdS}$ is larger than
the bubble radius $\bar R$. This can be directly seen
from the explicit bounce solution and it was interpreted \cite{cdl}
as the gravitational quenching of the tunneling at
$\bar R$ large enough.
Therefore we can not expect to have the standard four dimensional
gravity at initial stages of Minkowski evolution of the
the configuration considered. However at later stages
the size of the "Minkowski" bubble grows over
the $R_{AdS}$ and then we expect proper Newton law
in four dimensions at the intermediate distances.
We would like to conjecture  possible cosmological
implications of the fact that the  early  Universe
was actually five dimensional object. Note that in terms
of four dimensional gravity $T_{RS}$ can be interpreted
as the bare cosmological constant
\beq
\label{bar}
T_{RS}=\frac{\Lambda_4}{k_4}
\eeq
In assumption of the localization of gravity
$k_4=\frac{k_5}{R_{AdS}}$. Hence in view of Eq.(\ref{eff}),(\ref{bar})
the effective cosmological constant
is less than the bare one provided
the charge of the R brane produced is sufficiently small.

Let us mention the inspiring relation between the UV and IR scales
is such
world. Since the coordinate in the fifth dimension usually is identified
with the renormalization group scale in 4D world we see that the regulator
brane provides the effective UV cutoff which increases in time. On the
other hand the regulator brane is glued with our brane at the
asymptotics of 4D manifold. The size of our expanding universe
is fixed by the size of the internal brane expanding in the Minkowski
space. Since the size of the internal brane providing the IR cutoff
and the size of the regulator brane amounting to the UV cutoff
are related geometrically
\beq
R_{UV}=R_{IR}(1-\frac{T_{RS}^2}{4T_{R}^2})^{1/2}
\eeq
we have UV-IR relation at all times. This
relation reminds the similar relation in the noncommutative
QFT. Since noncommutativity can be attributed to the external
B field it can be treated just as UV-IR relation in the external
field. Here we see similar phenomena for the higher  rank fields.

5. Let us mention  a few generalizations of the
bounce solution above.

First, we can have multiple RS branes in the final state.
In that case we have additional pair of the R brane segments
of the same radius $\bar R$ for each new RS  brane.
The metric between
the n-th and n+1 -th RS branes reads
\beq
ds^{2}_{n}=\frac{dz^2+d\rho^2+\rho^2d\Omega^2_{3}}
{(1-\frac{(z-a_{n})^2+\rho^2}{R^{2}_{AdS}})^2}
\eeq
where $a_{n}=(2n-1)a$.
We choose z=0 at the center of the left R brane segment.
The induced cosmological constant is independent on n and
equals to
\beq
\Lambda_4=-\frac{12}{(R_{AdS}^2 -{a}^2)}.
\eeq
The distance
between branes in fifth dimension is $\delta x_{5}=2a$.
Let us interpret the RS branes as D branes which are neutral
with respect to the external NS field.
The picture described would amount to the
generic U(N) gauge group on their worldvolume.
Let us note  that since the distance between branes can be
interpreted as the vacuum expectation values of the scalar
field on the worldvolume of D branes  we have no
moduli associated to scalars in this solution.

Second we would like to describe the metric for the
induced brane production discussed in \cite{gs}.
The "fish"  type configuration consists of the
external brane and pair of R brane segments.
In that case we have flat metric inside and two pieces
of  spherical  metrics  sewed along the external brane.
The metric to the right reads
\beq
\label{plus1}
ds^{2}=\frac{dz^2+d\rho^2+\rho^2d\Omega^2_{3}}
{(1+\frac{(z+a)^2+\rho^2}{R^{2}_{S}})^2}
\eeq
while the metric to the left   reads
\beq
\label{minus1}
ds^{2}=\frac{dz^2+d\rho^2+\rho^2d\Omega^2_{3}}
{(1+\frac{(z-a)^2+\rho^2}{R^{2}_{S}})^2}
\eeq
The matching is like above with the appropriate
change of notations.

Note that the induced brane production is
relevant for all scenarios of RS
type in the external field.
The RS brane is neutral with respect to the
external field however it is unstable due
to the decay into the pair of
charged branes. Indeed one can use result of \cite{gs} to
get the probability
of the process with exponential accuracy. To avoid the
contradiction with the
cosmological data one has restrict the value of the external field.

Finally we can construct
explicit solution for the induced production
pair of R branes with RS brane inside. In this
case the four-brane junction is used.
If the tension of inducing brane is bigger than $T_{RS}$,
we have the "fish" type configuration with four pieces
of spherical metric  sewed. Otherwise, we have an
"apple" type configuration with
four pieces
of AdS  metric sewed.

6. In this letter we suggested new scenario for the brane world which
is free from the artificial branes with the negative tension
and jumps of the cosmological constant.
In
our scenario 4d world is localized on the brane which is
produced in the external higher-form field
via tunneling but is static with
respect to the higher dimensions.
As in other brane world constructions
we have a universal mechanism of decreasing of the
four dimensional cosmological constant. Some
feature of our model is that the
metric in our world corresponds to the
AdS space.
We would like to emphasize that in our model
early Universe is five dimensional hence one could expect
cosmological implications.

The work of A.G is partially supported by grants INTAS-99-1705 and
RFBR-98-01-00327
and the work of K.S by grant INTAS-97-0103.

\end{document}